%% file: main.tex
\documentclass[10pt,twocolumn,letterpaper]{article}
\usepackage{cvpr}              
\usepackage{svg}
\usepackage{graphicx}
\usepackage{multirow}
\usepackage[table]{xcolor}
\usepackage[accsupp]{axessibility}
\input{preamble}
\definecolor{cvprblue}{rgb}{0.21,0.49,0.74}
\usepackage[pagebackref,breaklinks,colorlinks,allcolors=cvprblue]{hyperref}

\title{Semantic Noise Reduction via Teacher-Guided Dual-Path Audio-Visual Representation Learning}

\author{
Linge Wang$^{1,2}$~~~~
Yingying Chen$^{1,2\dagger}$~~~~
Bingke Zhu$^{1,2}$~~~~
Lu Zhou$^{1,2}$~~~~
Jinqiao Wang$^{1,2,3\dagger}$\\
  $^{1}$~Foundation Model Research Center, Institute of Automation, \\ Chinese Academy of Sciences, Beijing, China \\ 
  $^{2}$~School of Artificial Intelligence, University of Chinese Academy of Sciences, Beijing, China\\
  $^{3}$~Objecteye Inc., Beijing, China\\
  {\tt\small  wanglinge2024@ia.ac.cn} \\
  {\tt\small \{yingying.chen,bingke.zhu,lu.zhou,jqwang\}@nlpr.ia.ac.cn}\\
  {\tt\small $^\dagger$Corresponding author} \\
}

\begin{document}
\maketitle
\begin{abstract}
\input{0_abstract}
\end{abstract}
\input{1_intro}
\input{2_related_work}
\input{3_method}

\input{4_Experiments}
\input{5_Conclusion}
\clearpage
\input{6_Acknowledgement}
{
    \small
    \bibliographystyle{ieeenat_fullname}
    \bibliography{main}
}


\end{document}

%% file: 0_abstract.tex
Recent advances in audio-visual representation learning have shown the value of combining contrastive alignment with masked reconstruction.  However, jointly optimizing these objectives in a single forward pass forces the contrastive branch to rely on randomly visible patches designed for reconstruction rather than cross-modal alignment, introducing semantic noise and optimization interference.
We propose TG-DP, a Teacher-Guided Dual-Path framework that decouples reconstruction and alignment into separate optimization paths. By disentangling the masking regimes of the two branches, TG-DP enables the contrastive pathway to use a visibility pattern better suited to cross-modal alignment. A teacher model further provides auxiliary guidance for organizing visible tokens in this branch, helping reduce interference and stabilize cross-modal representation learning.
TG-DP achieves state-of-the-art performance in zero-shot retrieval. On AudioSet, it improves R@1 from 35.2\% to 37.4\% for video-to-audio retrieval and from 27.9\% to 37.1\% for audio-to-video retrieval. The learned representations also remain semantically robust, achieving state-of-the-art linear-probe performance on AS20K and VGGSound. Taken together, our results suggest that decoupling multimodal objectives and introducing teacher-guided structure into the contrastive pathway provide an effective framework for improving large-scale audio-visual pretraining. Code is available at \url{https://github.com/wanglg20/TG-DP}.


%% file: 1_intro.tex
\section{Introduction}
\label{sec:intro}
\begin{figure}[t]
    \centering
    \includegraphics[width=\columnwidth]{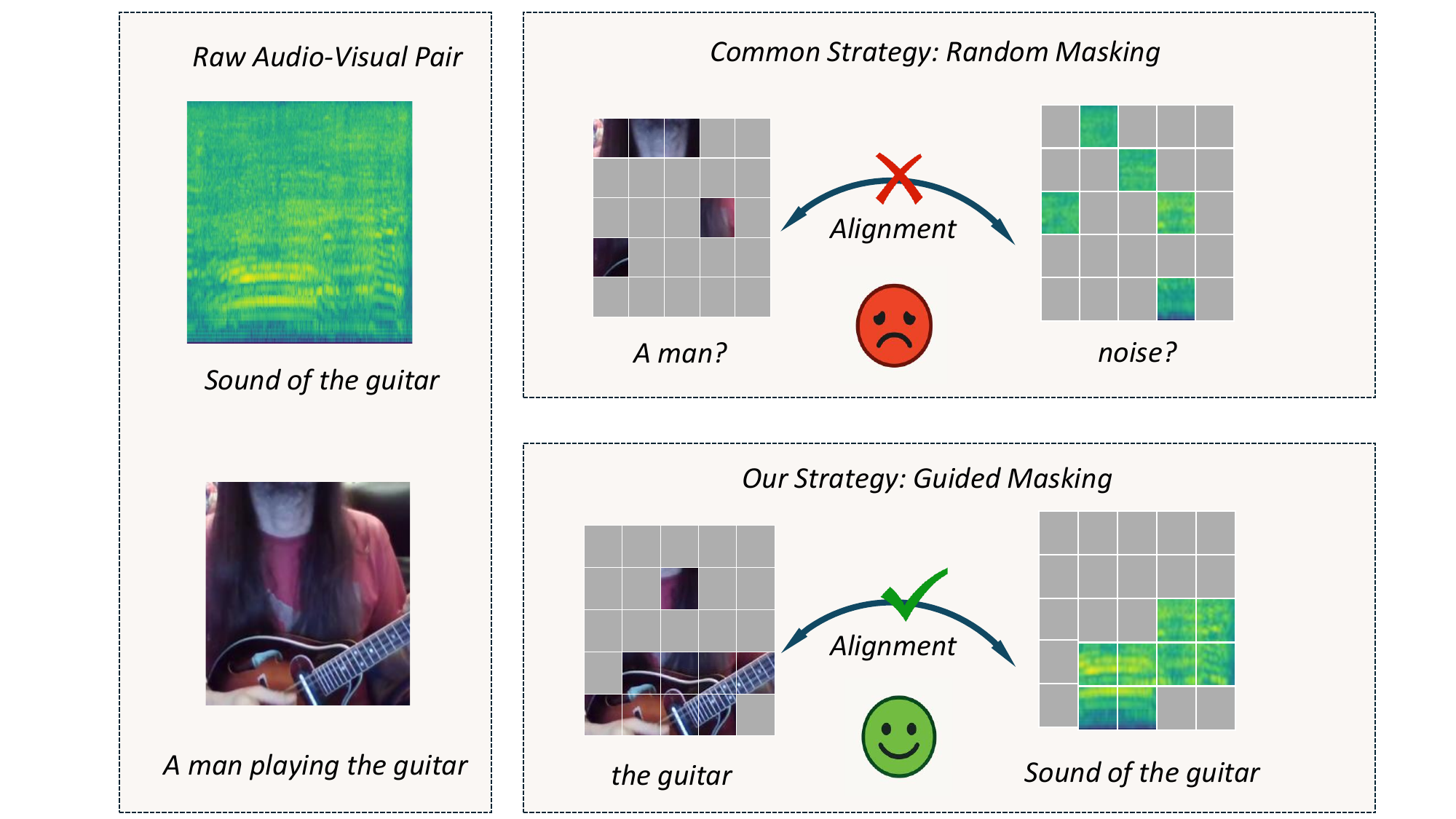}
    \caption{In existing contrastive masked autoencoder pretraining frameworks, random masking may expose patches with limited cross-modal relevance, such as silent spectrogram regions or visually uninformative background areas. Aggregating such partially observed tokens into global representations can introduce semantic noise into the contrastive branch and weaken cross-modal alignment. In contrast, our framework decouples the masking regimes of reconstruction and alignment, allowing the contrastive branch to use a lower masking ratio better suited to cross-modal alignment, with teacher guidance providing additional structure over the visible tokens.}
    \label{fig:motivation}
\end{figure}
Multimodal representation learning \cite{ngiam2011multimodal,baltruvsaitis2018multimodal,srivastava2012multimodal,tsai2019multimodal,alayrac2020avln,akbari2021vatt} has advanced rapidly with the emergence of large-scale self-supervised objectives. In the audio-visual domain, progress has been driven primarily by two complementary paradigms: masked autoencoding (MAE) \cite{he2022masked,georgescu2023audiovisual} for learning modality-specific structure through reconstruction, and contrastive learning (CL) \cite{oord2018representation,arandjelovic2017look,owens2018audio} for aligning heterogeneous modalities in a shared embedding space. Recent frameworks \cite{gong2022cavmae,araujo2025cav,huang2023mavil,guo2024crossmae} show that combining MAE and CL can produce strong audio-visual representations by leveraging both intra-modal reconstruction signals and cross-modal correspondence constraints.

Despite their complementarity, jointly optimizing MAE and CL within a shared framework still faces two central challenges. 
\textbf{First, semantic noise.} In prevailing approaches, the global tokens used for contrastive alignment are learned from randomly visible patches inherited from the masking process. Such visibility patterns are not designed for cross-modal matching and may preserve regions with limited semantic relevance, such as silent audio segments or visually uninformative backgrounds. As illustrated in Figure~\ref{fig:motivation}, this can contaminate the aggregated global representations and weaken fine-grained cross-modal alignment. 
\textbf{Second, optimization interference.} MAE and CL pursue inherently different learning objectives: MAE emphasizes reconstruction fidelity from partial observations, whereas CL encourages semantically invariant representations for cross-modal matching. When both objectives are imposed on shared token representations under a coupled training pathway, their gradient signals can interfere with each other and compromise effective optimization.

To address these issues, we propose \textbf{TG-DP}, a \textbf{T}eacher-\textbf{G}uided \textbf{D}ual-\textbf{P}ath framework that separates reconstruction and alignment into distinct optimization pathways while preserving the benefits of both objectives. Our framework consists of three key components. 
\textbf{(1) Dual-path decoupling.} We separate MAE reconstruction and contrastive alignment into two independent forward passes, each with its own masking regime and gradient flow. This design allows the contrastive branch to adopt a visibility pattern tailored to cross-modal alignment rather than being constrained by reconstruction-oriented random masking, thereby reducing optimization interference between the two objectives. 
\textbf{(2) Auxiliary teacher-student guidance.} A teacher encoder operating on full, unmasked inputs provides additional semantic guidance during training, supplying holistic context that complements the student's partially observed inputs. 
\textbf{(3) Structured masking for alignment.} Based on the teacher signals, the contrastive branch is encouraged to retain a more structured set of visible tokens, helping it focus on regions that are more useful for cross-modal correspondence and reducing the impact of noisy or weakly informative observations.

Extensive experiments on AudioSet-2M \cite{gemmeke2017audio} and VGGSound \cite{chen2020vggsound} show that TG-DP improves audio-visual retrieval while maintaining strong semantic robustness in the learned representations. As a result, our method achieves state-of-the-art performance on both classification and cross-modal retrieval benchmarks, suggesting that decoupling multimodal objectives and disentangling branch-specific masking strategies provide an effective recipe for large scale audio-visual pretraining.

In summary, our main contributions are as follows:
\begin{itemize}
\item
We introduce a dual-path training architecture that disentangles masked reconstruction and contrastive alignment into two separate forward passes. This design allows the two objectives to operate on distinct masked views, so that reconstruction and cross-modal alignment can each rely on visibility patterns better suited to their own learning goals.
\item
We propose a teacher-guided training mechanism that provides auxiliary semantic guidance and helps structure the visible tokens used by the contrastive branch, improving the stability of cross-modal representation learning.
\item
We achieve state-of-the-art results on large scale audio-visual classification and retrieval benchmarks, demonstrating the effectiveness and robustness of the proposed framework.
\end{itemize}

%% file: 2_related_work.tex
\section{Related Works}
\label{sec:Related Works}

\subsection{Audio-Visual Representation Learning}

Audio-visual representation learning leverages the natural co-occurrence of sound and visuals as a free supervisory signal. Early work such as SoundNet~\cite{aytar2016soundnet} used student–teacher distillation from pretrained image CNNs into an audio network, while correspondence-based methods like L3-Net~\cite{arandjelovic2017look} and ambient-sound supervision~\cite{owens2016ambient} learned joint features by predicting whether an image–audio pair originates from the same video. These simple co-occurrence objectives were shown to yield strong semantic features and even emergent localization capability~\cite{arandjelovic2018objects,zhao2018soundofpixels}.

With the rise of contrastive learning, many approaches adopted instance-discrimination objectives to align synchronized audio–video pairs~\cite{morgado2021crossmodal}. Extensions explored multi-scale contrast or stronger invariances to temporal transformations~\cite{kim2024equiav}, while also demonstrating benefits for localization and grounding~\cite{arandjelovic2018objects,chen2021localizing}. These methods established contrastive alignment as a core paradigm in audio-visual pretraining.

Motivated by the success of masked autoencoders, recent work integrates masked reconstruction with contrastive learning. AV-MAE~\cite{georgescu2023audiovisual} applies MAE to both modalities, while CAV-MAE~\cite{gong2022cavmae} combines MAE-style reconstruction with cross-modal contrastive objectives. Follow-ups such as MaViL~\cite{huang2023mavil}, CrossMAE~\cite{guo2024crossmae}, AVSiam~\cite{lin2024avsiam}, and VAB~\cite{su2024vab} refine masking strategies, fusion, and latent-space modeling. Broader multimodal frameworks—including ImageBind~\cite{girdhar2023imagebind}, LanguageBind~\cite{zhu2024languagebind}, and DenseAV~\cite{hamilton2024denseav}—extend alignment to additional modalities or emphasize dense, region-aware supervision.

Despite this progress, many contrastive audio-visual MAE frameworks still mix reconstruction and contrastive objectives within a shared token space, potentially creating competing optimization signals. Recent work such as CAV-MAE Sync~\cite{araujo2025cav} begins to address these limitations through finer temporal alignment. Our work further investigates how training frameworks and objective separation influence audio-visual correspondence and representation quality.

\subsection{Teacher-Student Learning}

Teacher-student learning provides a general framework for transferring knowledge between models, typically by guiding a student network with signals distilled from a stronger, smoother, or more stable teacher. The paradigm was popularized by knowledge distillation, where a student is trained to match the teacher’s soft predictions or intermediate representations~\cite{hinton2015distillation,romero2015fitnets}. Early variants explored compressing large image classifiers or transferring intermediate hints, showing that teacher-driven supervision can enrich gradients and stabilize training.

Self-distillation later emerged as a powerful form of teacher-student learning. In temporal-ensemble methods, the teacher is an exponential moving average (EMA) of the student, producing stable targets over training time. BYOL~\cite{grill2020byol} demonstrated that self-distillation without negative samples can learn strong visual representations, while SimSiam~\cite{chen2021simsiam} further simplified the design by constraining the student network with stop-gradient operations. Building on these ideas, DINO~\cite{caron2021emerging} introduced a ViT-based framework where a teacher generates supervisory signals for a student, yielding both improved global embeddings and emergent object localization capabilities. These works established EMA-based self-distillation as a robust and general recipe for representation learning.

In multimodal contexts, teacher-student learning has been used to fuse information across modalities or to enhance one modality with guidance from another. SoundNet~\cite{aytar2016soundnet} distilled visual knowledge into an audio network using millions of noisy videos, demonstrating that cross-modal distillation can bootstrap underrepresented modalities. CLIP-style cross-modal training has further inspired using teacher signals to stabilize multimodal contrastive learning, where teacher encoders provide smoother embeddings or alignment constraints~\cite{radford2021clip, alayrac2020avln}. Overall, teacher-student frameworks have evolved from model compression to a central principle for scalable representation learning. Their stability, smoothing effects, and ability to encode high-level semantic priors make them particularly effective in audio-visual learning, where cross-modal differences and temporal misalignment pose significant challenges. Our work adopts this perspective by designing a teacher-guided training pipeline that decouples objectives and improves audio-video alignment at both global and token levels.

Building on these ideas, recent self-supervised approaches such as
DINO~\cite{caron2021emerging}, Data2Vec~\cite{baevski2022data2vec}, and
iBOT~\cite{zhou2022ibot} show that teacher–student consistency is a strong
implicit regularizer, stabilizing training, suppressing noise, and promoting
high-level semantic structure. This consistency is particularly valuable in
multimodal learning, where modalities differ in noise, temporal resolution,
and semantic density. 

%% file: 3_method.tex
\section{Method}
\subsection{Overview}
\begin{figure*}[t]
    \centering
    \includegraphics[width=\textwidth]{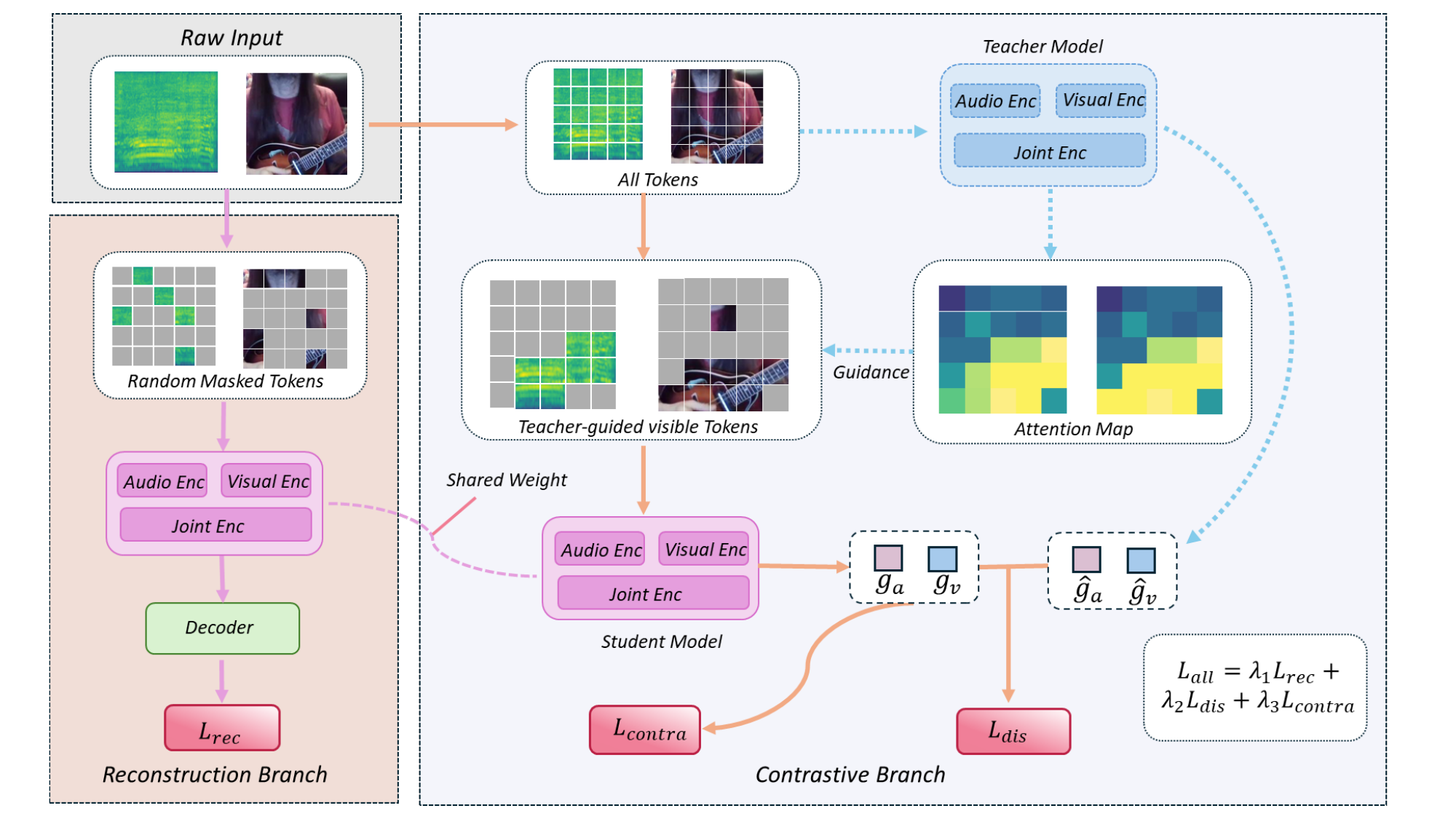}
    \caption{Overall pipeline of our proposed framework.
    The model consists of two objective-specific forward passes:
    (1) the \textbf{reconstruction branch} (left) learns modality-specific generative representations by randomly masking both audio and visual patches and reconstructing them through separate decoders.
    (2) the \textbf{contrastive branch} (right) adopts a teacher-student scheme for cross-modal alignment under a lower masking ratio.
    The teacher model observes full, unmasked inputs and produces attention-based token priority cues to guide the student's masking strategy.
    The student encodes the masked tokens, appends modality-specific global tokens $(g_v, g_a)$, and aligns them via the InfoNCE-based contrastive loss $L_{\text{contra}}$.
    An additional distillation loss $L_{\text{dis}}$ encourages consistency between teacher and student global embeddings.
    The encoders in the reconstruction branch share weights with the student encoders in the contrastive branch.
    The final training objective combines all terms as
    $L_{\text{all}} = \lambda_1 L_{\text{rec}} + \lambda_2 L_{\text{dis}} + \lambda_3 L_{\text{contra}}$.
    }
    \label{fig:pipeline}
\end{figure*}

Figure~\ref{fig:pipeline} illustrates our overall framework, which separates masked reconstruction and cross-modal alignment into two objective-specific forward passes. In the \emph{reconstruction branch}, randomly masked audio-visual tokens are reconstructed, encouraging the encoder to learn strong modality-specific representations under heavy masking. In the \emph{contrastive branch}, we adopt a teacher-student design with a lower masking ratio. The teacher processes full, unmasked inputs and produces token-priority cues and global embeddings that provide full view semantic guidance. These cues bias the student's masking toward potentially informative regions, yielding a more suitable view for contrastive learning. The student then encodes the masked tokens, appends global tokens, and aligns the resulting audio and visual global embeddings via a cross-modal contrastive loss, while an additional distillation loss encourages the student globals to remain consistent with the teacher's full view representation. The two branches are jointly optimized during training and together form the basis of our method.
\subsection{Backbone: CAV-MAE Sync}
We first briefly summarize the CAV-MAE Sync backbone that we adopt in our framework.
Given an input video with its corresponding audio, 
we sample $T$ frames from the video and extract log-Mel spectrogram segments aligned with each frame.
During training, a single frame and its temporally aligned audio segment are randomly selected as one training pair. 
Both RGB and spectrogram inputs are patchified into tokens, where a learnable global token and a set of register tokens are inserted at the beginning of each modality sequence.
A random subset of patches is masked prior to encoding for the reconstruction objective.
The visible tokens of each modality are then encoded separately by $E_v$ and $E_a$:
\begin{align}
Z^{v} = E_v(X^{v}), \qquad Z^{a} = E_a(X^{a}).
\end{align}
Visible audio and visual tokens are concatenated and passed through the joint encoder–decoder $(J, D)$ 
to reconstruct the masked patches:
\begin{align}
\hat{M}^{v}, \hat{M}^{a} = D(J(\text{concat}(\text{vis}(Z^{v}), \text{vis}(Z^{a})))).
\end{align}
The reconstruction loss for each modality is defined as
\begin{align}
L_{\text{rec}}^{m}
= \frac{1}{|M_m|} \sum_{i \in M_m}
\| \hat{m}_i^{m} - m_i^{m} \|_2^2, \quad m \in \{v,a\}.
\end{align}

For cross-modal alignment, the global tokens serve as compact video level and audio level descriptors.
Let $g^{v}_i$ and $g^{a}_i$ denote the global tokens of sample~$i$ after their respective encoders and the joint layer.
A contrastive InfoNCE loss is then applied:
\begin{align}
L_{\text{contra}}
= - \frac{1}{N} \sum_{i=1}^{N}
\log \frac{\exp(\langle g^{v}_i, g^{a}_i \rangle / \tau)}
{\sum_{j=1}^{N} \exp(\langle g^{v}_i, g^{a}_j \rangle / \tau)}.
\end{align}
The overall training objective combines both reconstruction and contrastive terms:
\begin{align}
L_{\text{CAV-Sync}}
= \lambda_{\text{rec}} L_{\text{rec}}
+ \lambda_{\text{con}} L_{\text{contra}}.
\end{align}

\subsection{TG-DP Framework}

While CAV-MAE Sync introduces separate global and register tokens to reduce the burden on patch tokens, its contrastive objective is still optimized on representations that are jointly shaped by masked reconstruction. Under random masking and full frame reconstruction, the encoder must preserve substantial modality specific and weakly shared content, such as silent spectrogram regions, static backgrounds, or off-screen sources. When the same masked view is also used for contrastive learning, these factors can interfere with the formation of global embeddings for cross-modal alignment. As a result, the contrastive branch is not trained on a view that is optimally matched to semantic correspondence learning.

To address this issue, we redesign the training pipeline to (i) separate reconstruction and contrastive learning into two objective-specific forward passes with different masking regimes, and (ii) introduce a teacher-student guided masking strategy that provides structured semantic guidance for the contrastive branch.


\paragraph{Dual-pass training.}
To address the issues described above, we reformulate pretraining into two objective specific forward passes for each audio-visual pair, as illustrated in Figure~\ref{fig:pipeline}. 
Each sample is processed twice under different masking configurations: one pass for the \textit{reconstruction branch} and the other for the \textit{contrastive branch}. 
The two branches share the same encoders and joint layers, but each loss is computed from its own masked view: only the reconstruction branch contributes to $L_{\text{rec}}$, while only the contrastive branch contributes to $L_{\text{contra}}$. 
As a result, masked reconstruction and cross-modal alignment are no longer optimized on the same token representation, which reduces the coupling between generative and discriminative objectives.

In the reconstruction branch, a large masking ratio (typically 75\%) is used to encourage the model to infer missing content from limited context. 
In contrast, the contrastive branch adopts a smaller masking ratio (50\% in our setting), exposing more visible patches and preserving a more semantically complete context for global representation learning. 
This asymmetric masking provides a view better matched to contrastive alignment, while still retaining partial masking as a useful regularizer. 
Empirically, we find that this design leads to more stable optimization and stronger cross-modal alignment than applying both objectives on a shared masked view.

\paragraph{Teacher-student integration.}
As an auxiliary regularization for the contrastive branch, we incorporate a lightweight teacher-student framework inspired by self-distillation.
The teacher network receives full, unmasked inputs from both modalities and produces global representations $[\hat{g}_{v}, \hat{g}_{a}]$ that provide full view semantic targets.
The student network, operating on the masked inputs, outputs its own global tokens $[g_{v}, g_{a}]$.
In addition to the contrastive loss, we apply a distillation loss that encourages the student globals to remain consistent with the teacher representations:
\begin{align}
L_{\text{dis}} =
\| g_{v} - \hat{g}_{v} \|_2^2 +
\| g_{a} - \hat{g}_{a} \|_2^2.
\end{align}
This mean-squared error term serves as a semantic anchoring signal for masked-view global embeddings, while the teacher parameters are updated by an exponential moving average (EMA) of the student parameters for temporal stability.
The final training objective combines reconstruction, distillation, and contrastive losses:
\begin{align}
L_{\text{all}}
= \lambda_{1} L_{\text{rec}}
+ \lambda_{2} L_{\text{dis}}
+ \lambda_{3} L_{\text{contra}}.
\end{align}
Here, $\lambda_{1}$, $\lambda_{2}$, and $\lambda_{3}$ are fixed hyperparameters that balance the three objectives. In our framework, reconstruction provides modality-specific generative supervision, distillation offers an auxiliary full view consistency constraint, and contrastive learning remains the primary objective for cross-modal alignment.

\paragraph{Teacher-guided masking strategy.}
To provide additional structure for the contrastive branch, we introduce a teacher-guided masking strategy when constructing the student view.
Specifically, we extract attention weights from the teacher's joint encoder, measuring how strongly each patch token interacts with the global token within each modality.
These scores are normalized across spatial tokens and used as token-priority cues for masking.
The student then retains the top-$k$ tokens with the highest scores as visible inputs, where $k$ is determined by the masking ratio of the contrastive branch.
Rather than relying purely on random masking, this deterministic selection biases the student view toward regions that are more likely to be informative under the teacher's full view representation, yielding a more structured input for contrastive learning.
In this way, teacher-guided masking provides a simple mechanism for injecting teacher-derived semantic bias into the contrastive branch.

%% file: 4_Experiments.tex
\section{Experiments}
\subsection{Dataset}
\label{ssec:dataset}

\noindent\textbf{AudioSet-2M}. AudioSet-2M \cite{gemmeke2017audio} is a large-scale audio-visual dataset originally curated for audio event classification, consisting of approximately two million 10-second video-audio clips collected from YouTube and annotated with 527 sound event classes. Despite its wide coverage and diversity, the dataset was not explicitly designed for audio-visual learning. As a result, a considerable portion of the clips contain loosely correlated or even mismatched visual-audio content (e.g., off-screen sound sources or silent scenes), introducing substantial noise into cross-modal alignment. Moreover, because all samples are hosted on YouTube, practical access is affected by video availability and regional restrictions. In our implementation, approximately 350,000 samples could not be obtained due to unavailable or inaccessible YouTube videos. Among the downloaded data, we further excluded roughly 50,000 videos with missing audio tracks or corrupted frames, resulting in 1,390,395 valid video-audio pairs for pre-training. For downstream linear probing, we adopt the balanced subset AudioSet-20K, while for retrieval evaluation, we follow the official split provided by \cite{gong2022cavmae}.

\noindent\textbf{VGGSound}. VGGSound \cite{chen2020vggsound} contains 200,000 10-second YouTube video-audio clips covering 309 sound event classes. In contrast to AudioSet, VGGSound was explicitly designed for audio–visual learning, ensuring that each video contains a visible sound source corresponding to the audio track. This design makes VGGSound widely regarded as having high-quality cross-modal alignment, with a significantly lower proportion of irrelevant or mismatched visual–audio pairs. In our experiments, we use the VGGSound training split for linear probing, and adopt the retrieval split provided by \cite{araujo2025cav} for zero-shot audio–visual retrieval evaluation.

\subsection{Main Result}
\input{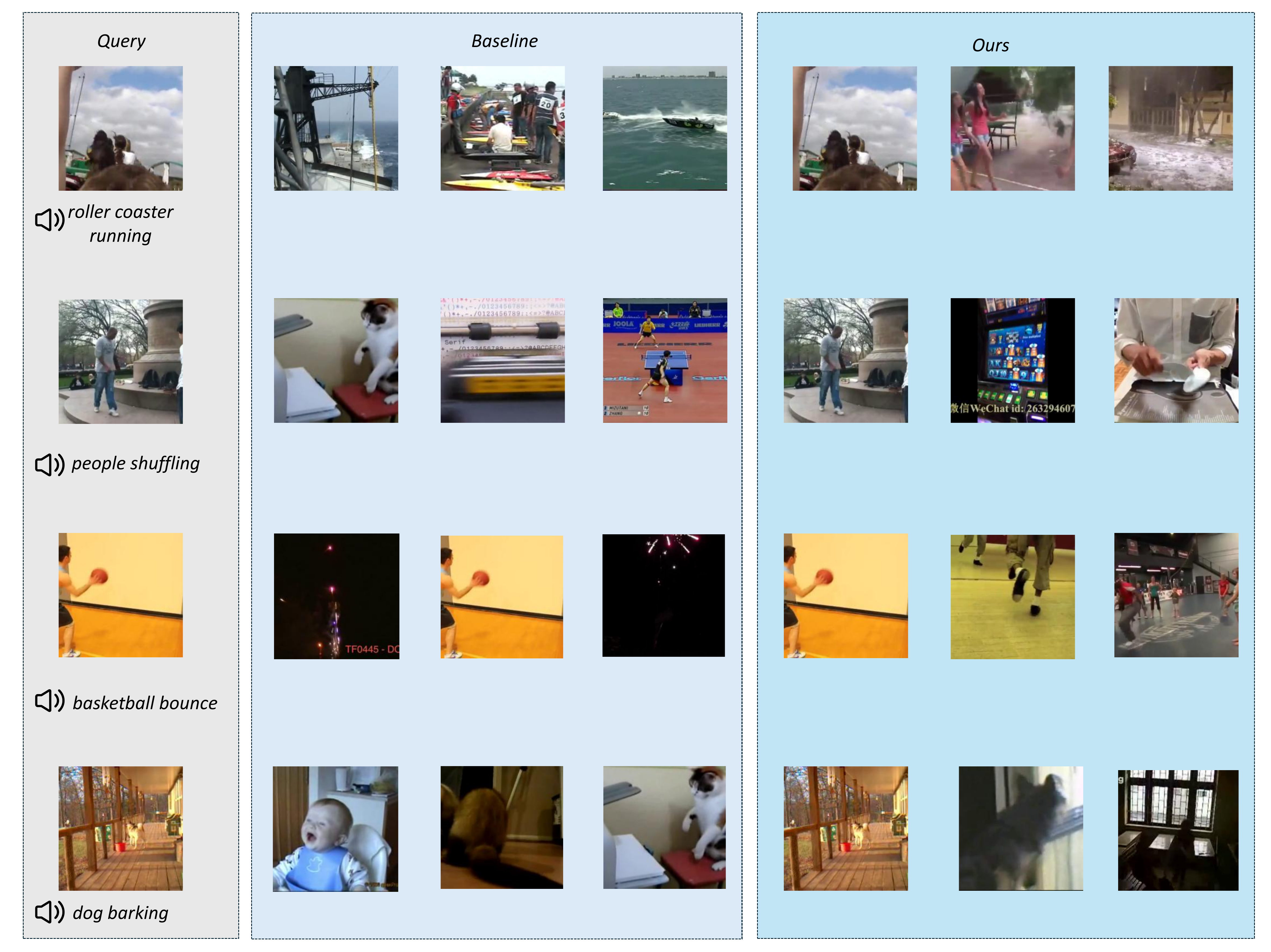}
\input{tables/classification}
\input{tables/unimodal}

\paragraph{Zero-shot Audio-Visual Retrieval.}
We evaluate cross-modal alignment quality through zero-shot audio-visual retrieval. Following~\cite{araujo2025cav}, retrieval is performed by computing cosine similarity between audio and visual global tokens, and ranking candidates accordingly. We report Recall@1, Recall@5, and Recall@10 for both retrieval directions (Visual$\rightarrow$Audio and Audio$\rightarrow$Visual) on the official evaluation splits of AudioSet and VGGSound. The results are summarized in Table~\ref{tab:retrieval}.

As shown in Table~\ref{tab:retrieval}, our method achieves the best zero-shot retrieval performance across both datasets and both retrieval directions among the directly comparable training-free methods. Compared with CAV-MAE Sync~\cite{araujo2025cav}, our approach consistently improves recall at all retrieval depths, demonstrating the effectiveness of the proposed training framework for cross-modal representation learning. For reference, we also report the gray-highlighted VAB-Encodec result as a fine-tuned reference, since it requires task-specific adaptation on the target retrieval dataset and is therefore not directly comparable to the training-free retrieval results of the other methods.

We further observe that the gain is particularly pronounced in the Audio$\rightarrow$Visual direction, which is typically the more challenging setting. We attribute this asymmetry to two factors in existing audio-visual pretraining frameworks. First, many methods initialize both encoders from vision-domain pretrained weights. While this initialization is naturally suitable for the visual branch, it is less well matched to the statistical structure of audio spectrograms, making the audio encoder more vulnerable when reconstruction and contrastive learning are optimized jointly. Second, audio semantics are often distributed more sparsely across tokens and are less redundant than visual semantics, so heavy masking is more likely to remove globally important cues on the audio side. Our framework alleviates these issues by decoupling the two objectives and, in particular, using a lower masking ratio in the contrastive branch to preserve a more semantically complete audio view. As a result, the improvement is especially clear in Audio$\rightarrow$Visual retrieval.

\paragraph{Audio-Visual Classification.}
We evaluate the representation quality of our pretrained model through \emph{attention probing}, following the evaluation protocol of \cite{araujo2025cav}. Specifically, for each video sample $i$ of length $T$, we first extract the global tokens $g_v$ and $g_a$ from the visual and audio encoders at each temporal step $t$, and concatenate them to obtain a unified sequence $C_i$ of length $T$. A learnable [CLS] token is prepended to this sequence to aggregate temporal context, forming the final sequence $C'_i=\{\text{CLS}, C_{i,1}, \ldots, C_{i,T}\}$, which is then fed into a lightweight classification head $f_{\text{cls}}$ composed of a two-layer transformer followed by a linear projection. The prediction is given by $\hat{y}_i=f_{\text{cls}}(C'_i)$. Although this protocol was referred to as \emph{linear probing} in CAV-MAE Sync~\cite{araujo2025cav}, we use the term \emph{attention probing} here because the classifier includes attention layers.

For multi-label classification on AudioSet, we adopt binary cross-entropy loss, while for single label classification on VGGSound, we use cross entropy loss. In both cases, the pretrained encoders are frozen and only the lightweight classification head is optimized. For AudioSet-20K, the classifier is trained on the official training split and evaluated on the balanced evaluation subset using mean Average Precision (mAP). For VGGSound, the classifier is trained on the official training split and evaluated on the test set using Top-1 accuracy.

As shown in Table~\ref{tab:linearprobing}, our model achieves strong and consistent performance under this frozen encoder protocol across both datasets. On AS20K, it reaches an mAP of $32.0$, outperforming existing approaches pretrained on AudioSet-2M. On VGGSound, it achieves $52.7\%$ Top-1 accuracy, matching or exceeding prior results under the same protocol. Notably, our model is pretrained on only a 1.4M subset of AudioSet-2M, yet still transfers effectively to downstream classification. We further evaluate unimodal transfer on AS20K in Table~\ref{tab:unimodal}. Compared with CAV-MAE Sync, our model improves both audio-only mAP (29.3 $\rightarrow$ 31.2) and vision-only mAP (14.3 $\rightarrow$ 17.8), suggesting that the learned representations are strengthened not only at the cross-modal level but also within each individual modality.

\begin{table}[h]
\centering
\setlength{\tabcolsep}{10pt}
\begin{tabular}{lcc}
\toprule
\textbf{Model} & \textbf{Per-epoch (s)} & \textbf{Total } \\
\midrule
baseline & 730  & 7.1 h \\
Ours          & 1045 & 10.2 h \\
\bottomrule
\end{tabular}
\caption{Pretraining runtime comparison. Our method introduces moderate additional training cost due to the extra forward pass and EMA teacher, but incurs no extra computation at inference time.}
\label{tab:time-comparison}
\end{table}
\input{tables/ablation}

\subsection{Computation Cost}
Table~\ref{tab:time-comparison} compares the pretraining runtime.
Our method increases the per-epoch time from 730\,s to 1045\,s, leading to a total pretraining time of 10.2\,h versus 7.1\,h for the baseline.
The overhead comes from the additional training-time forward pass and the EMA teacher.
Importantly, these components are discarded after pretraining, so our method incurs no extra parameters or computation at inference time.

\subsection{Ablation}

\paragraph{Effect of Dual Forward Passes.}
We first examine the effect of separating reconstruction and contrastive learning into two objective specific forward passes.
In this ablation, we introduce an additional forward pass on top of the baseline single-pass training scheme, so that the reconstruction branch and contrastive branch are optimized from different views.
To isolate the effect of the dual-pass structure itself, both passes use the same masking ratio ($75\%$) as the baseline, while keeping all other training settings unchanged.

As shown in Table~\ref{tab:ablation_dualpass}, the dual-pass scheme brings moderate gains on VGGSound retrieval, especially in the more challenging Audio$\rightarrow$Visual direction (e.g., R@10: 58.1 $\rightarrow$ 60.1), while maintaining comparable performance on AS20K. These results suggest that separating the two objectives into different forward passes is beneficial, but its more important role is to provide the structural basis for applying different masking regimes to reconstruction and contrastive learning, which is explored in the following ablations.

\paragraph{Effect of Mask Ratio in the Contrastive Pass.}
We evaluate how the masking ratio in the contrastive forward pass affects cross-modal alignment.
While the reconstruction branch follows the MAE convention with a fixed high masking ratio, the contrastive branch requires a different trade-off: lower masking preserves more semantic context for alignment, whereas higher masking provides stronger regularization and lower computation.
To isolate this factor, we vary the contrastive masking ratio over \{0.00, 0.20, 0.50, 0.65, 0.75\}, while keeping the reconstruction branch unchanged.

As shown in Table~\ref{tab:ablation_maskratio}, very low masking ratios (0.00 and 0.20) achieve the strongest retrieval results on VGGSound, indicating that cross-modal alignment benefits from more semantically complete views.
However, these settings lead to noticeably worse AS20K classification performance, suggesting that removing masking almost entirely weakens the regularization effect of the contrastive branch.
At the other extreme, higher masking ratios (0.65 and 0.75) substantially degrade retrieval, especially in the Audio$\rightarrow$Visual direction, where sparse audio semantics are more easily damaged under heavy masking.
Among all settings, a masking ratio of 0.50 provides the best overall balance: it maintains strong retrieval performance, achieves the best AS20K classification accuracy ($32.0$ mAP), and avoids the additional cost of near-full contrastive views.
We therefore adopt 0.50 as the default masking ratio in the contrastive branch.
These results suggest that the contrastive branch should not follow the same high masking regime as reconstruction. A lower masking ratio is more suitable, as it preserves richer global semantic context while still providing useful regularization.

\paragraph{Effect of Teacher–Student Distillation.}
We further examine the effect of adding a teacher–student distillation loss in the contrastive branch.
In this ablation, the student is trained either with or without an auxiliary loss that encourages its global representations to match those generated by a teacher network.
All other training settings are kept identical to isolate the contribution of the distillation objective.

As shown in Table~\ref{tab:ablation_teacher}, incorporating distillation produces minor but consistent gains across several metrics.
In particular, we observe improvements in the more challenging A→V retrieval direction (R@10:63.5 → 64.3) and in AS20K classification (30.5 → 32.0 mAP), suggesting that teacher supervision helps stabilize the student’s global representations.
However, the improvements on V→A retrieval are marginal, indicating that the benefit of distillation is relatively modest compared to other components in our framework.
Overall, distillation provides a lightweight enhancement to representation quality, but is not a dominant factor in the final performance of our model.

\paragraph{Effect of Teacher-Guided Masking.}
We evaluate how different strategies for selecting visible tokens in the contrastive branch affect performance.
Table~\ref{tab:ablation_guidedmask} compares three masking schemes:
(1) \textbf{Random Masking}, where visible tokens are sampled uniformly at random;
(2) \textbf{Distinct Guided Mask}, where we directly select the top-$k$ tokens according to the teacher attention scores; and
(3) \textbf{Probabilistic Guided Mask}, where Gumbel noise is added to the teacher scores before top-$k$ selection, introducing stochasticity into the token ranking.

As shown in Table~\ref{tab:ablation_guidedmask}, the Distinct Guided Mask performs competitively and is generally more stable than random masking across retrieval and classification metrics.
This suggests that teacher-based token ranking provides a useful structured prior for constructing the contrastive view.
In contrast, the Probabilistic Guided Mask consistently underperforms the deterministic variant and in some cases falls below the random baseline.
A likely reason is that adding Gumbel noise perturbs the token ranking and makes the visible set less consistent across samples, which weakens the structured guidance provided to the contrastive branch.
Overall, these results suggest that deterministic teacher-guided masking is a more effective way to construct the contrastive view than noisy ranking-based selection.

%% file: tables/retrieval.tex
\begin{table*}[!ht]
\centering
\setlength{\tabcolsep}{5.2pt}
\resizebox{\linewidth}{!}{
\begin{tabular}{@{}l ccc ccc c ccc ccc@{}}
\toprule
& \multicolumn{6}{c}{\textbf{Vision $\rightarrow$ Audio}} & & \multicolumn{6}{c}{\textbf{Audio $\rightarrow$ Vision}}\\
\cmidrule(lr){2-7}\cmidrule(lr){9-14}
& \multicolumn{3}{c}{AudioSet Eval Subset} & \multicolumn{3}{c}{VGGSound Eval Subset} & & \multicolumn{3}{c}{AudioSet Eval Subset} & \multicolumn{3}{c}{VGGSound Eval Subset}\\
\cmidrule(lr){2-4}\cmidrule(lr){5-7}\cmidrule(lr){9-11}\cmidrule(lr){12-14}
\textbf{Method} & R@1 & R@5 & R@10 & R@1 & R@5 & R@10 & & R@1 & R@5 & R@10 & R@1 & R@5 & R@10\\
\midrule
\textcolor{lightgray}{VAB-Encodec \cite{su2024vab}} & \textcolor{lightgray}{39.5} & \textcolor{lightgray}{65.4} & \textcolor{lightgray}{74.6} & \textcolor{lightgray}{33.5} & \textcolor{lightgray}{63.3} & \textcolor{lightgray}{74.3} & & \textcolor{lightgray}{37.5} & \textcolor{lightgray}{64.0} & \textcolor{lightgray}{73.7} & \textcolor{lightgray}{34.9} & \textcolor{lightgray}{62.7} & \textcolor{lightgray}{73.1}\\
\midrule
CAV-MAE \cite{gong2022cavmae} & 16.1 & 38.6 & 49.3 & 14.7 & 35.3 & 45.9 & & 9.5 & 22.6 & 32.4 & 8.3 & 23.8 & 32.4\\
CAV-MAE\textsuperscript{Scale+++} \cite{gong2022cavmae} & 18.8 & 39.5 & 50.1 & 14.8 & 34.2 & 44.0 & & 15.1 & 34.0 & 43.0 & 12.8 & 30.4 & 40.3\\
LanguageBind \cite{zhu2024languagebind} & 6.4 & 20.2 & 28.3 & 10.3 & 30.1 & 39.7 & & 4.4 & 15.0 & 22.5 & 6.5 & 22.7 & 33.5\\
AVSiam \cite{lin2024avsiam} & 19.7 & -- & -- & 19.0 & -- & -- & & 17.6 & -- & -- & 20.4 & -- & --\\
ImageBind \cite{girdhar2023imagebind} & 22.1 & 43.2 & 52.6 & 21.6 & 43.4 & 52.9 & & 20.8 & 42.6 & 51.6 & 20.7 & 43.2 & 53.4\\
DenseAV \cite{hamilton2024denseav}  & 26.3 & 49.9 & 58.5 & 24.2 & 48.2 & 58.3 & & 20.8 & 42.6 & 51.6 &  25.1 & 50.1 & 59.1\\

\addlinespace
CAV\mbox{-}MAE Sync \cite{araujo2025cav} & 35.2 & 58.3 & 67.6 & 27.9 & 51.7 & 61.8 & & 27.9 & 52.4 & 62.2 & 23.2 & 46.2 & 58.1\\
\midrule
\rowcolor{gray!15} \textbf{Ours} & \textbf{37.4} & \textbf{59.2} & \textbf{68.1} & \textbf{31.3} & \textbf{54.4} & \textbf{63.9} & & \textbf{37.1} & \textbf{59.7} & \textbf{68.2} & \textbf{30.3} & \textbf{53.1} & \textbf{64.2}\\
\bottomrule
\end{tabular}
}
\caption{
Zero-shot retrieval results on AudioSet and VGGSound for Vision$\rightarrow$Audio (V$\rightarrow$A) and Audio$\rightarrow$Vision (A$\rightarrow$V) tasks. 
Our model achieves \textbf{state-of-the-art zero-shot performance} across all retrieval metrics (R@1, R@5, R@10) on both datasets, outperforming existing \textbf{training-free} baselines such as ImageBind, AVSiam, and CAV-MAE variants without any fine-tuning.
For reference, we also report the result of VAB-Encodec, which is gray-highlighted because it relies on task-specific fine-tuning on the target retrieval dataset, unlike the other methods in this table that are evaluated in a training-free manner using frozen embeddings.
}\label{tab:retrieval}
\end{table*}

%% file: tables/classification.tex
\begin{table}[!ht]
\centering
\setlength{\tabcolsep}{2pt}       
\renewcommand{\arraystretch}{1.1} 
\small                            
\begin{tabular}{@{}lcccc@{}}      
\toprule
\textbf{Method} & \textbf{Pre.} & 
\textbf{AS20K(mAP)} & 
\textbf{VGG(Acc)} \\
\midrule
\textcolor{lightgray}{VAB-Encodec \cite{su2024vab}} &
\textcolor{lightgray}{AS2M+VGG} &
\textcolor{lightgray}{33.3} &
\textcolor{lightgray}{57.6} \\
\midrule
CAV-MAE \cite{gong2022cavmae}            & AS2M & 27.3 & --   \\
CAV-MAE\textsuperscript{Scale+++} \cite{gong2022cavmae} & AS2M & 25.3 & 51.6 \\
MaViL \cite{huang2023mavil}              & AS2M & 30.0 & --   \\
CAV-MAE Sync                              & AS2M & 30.5 & 52.7 \\
\midrule
\rowcolor{gray!12}
\textbf{Ours}                             & AS2M & \textbf{32.0} & \textbf{52.7} \\
\bottomrule
\end{tabular}
\caption{
Audio-visual classification results on AS20K (mAP) and VGGSound (Top-1 accuracy) using attention probing.
All models use frozen encoders.
}
\label{tab:linearprobing}
\end{table}

%% file: tables/unimodal.tex
\begin{table}[!ht]
\centering
\setlength{\tabcolsep}{6.5pt}
\resizebox{\linewidth}{!}{
\begin{tabular}{l cc}
\toprule
\textbf{Method} & \textbf{AS20K (Audio-only) mAP} & \textbf{AS20K (Vision-only) mAP} \\
\midrule
CAV-MAE Sync & 29.3 & 14.3 \\
Ours (TG-DP) & \textbf{31.2} & \textbf{17.8} \\
\bottomrule
\end{tabular}
}

\caption{Unimodal classification on AS20K under the same frozen-encoder attention probing protocol. TG-DP improves both audio-only and vision-only performance over CAV-MAE Sync, indicating that the learned representations are stronger not only for cross-modal alignment but also for single-modality transfer.}
\label{tab:unimodal}
\end{table}

%% file: tables/ablation.tex
\begin{table}[!ht]
\centering
\setlength{\tabcolsep}{6.5pt}
\resizebox{\linewidth}{!}{
\begin{tabular}{l ccc ccc c}
\toprule
\multirow{2}{*}{\textbf{Dual Pass}} 
& \multicolumn{3}{c}{\textbf{VGGSound (V$\rightarrow$A)}} 
& \multicolumn{3}{c}{\textbf{VGGSound (A$\rightarrow$V)}} 
& \textbf{Classification} \\
\cmidrule(lr){2-4} \cmidrule(lr){5-7} \cmidrule(lr){8-8}
& R@1 & R@5 & R@10 & R@1 & R@5 & R@10 & AS20K (mAP) \\
\midrule
w/o & \textbf{27.9} & 51.7 & 61.8 & 23.2 & 46.2 & 58.1 & 30.5 \\
\rowcolor{gray!15}
w/  & 27.1 & \textbf{51.8} & 61.2 & \textbf{27.4} & \textbf{50.7} & \textbf{60.1} & 30.4 \\
\bottomrule
\end{tabular}
}
\caption{
Ablation on the \emph{dual-pass} training scheme.
}
\label{tab:ablation_dualpass}
\end{table}

\begin{table}[!ht]
\centering
\setlength{\tabcolsep}{6.5pt}
\resizebox{\linewidth}{!}{
\begin{tabular}{l ccc ccc c}
\toprule
\textbf{Mask Ratio} 
& \multicolumn{3}{c}{\textbf{VGGSound (V$\rightarrow$A)}} 
& \multicolumn{3}{c}{\textbf{VGGSound (A$\rightarrow$V)}} 
& \textbf{Classification} \\
\cmidrule(lr){2-4} \cmidrule(lr){5-7} \cmidrule(lr){8-8}
& R@1 & R@5 & R@10 & R@1 & R@5 & R@10 & AS20K (mAP) \\
\midrule
0.00 & 30.4 & \textbf{56.0} & \textbf{64.2} & 29.8 & \textbf{55.6} & 64.5 & 29.6\\
0.20 & \textbf{32.5} & 55.7 & 64.1 & \textbf{31.7} & 54.5 & \textbf{64.8} & 30.4 \\
\rowcolor{gray!15}
0.50 & 31.3 & 54.4 & 63.9 & 30.3 & 53.1 & 64.3 & \textbf{32.0} \\
0.65 & 26.4 & 48.1 & 57.0 & 25.1 & 47.0 & 58.4 & 30.5 \\
0.75 & 27.1 & 50.4 & 60.1 & 26.2 & 49.4 & 60.8 & 30.7 \\
\bottomrule
\end{tabular}
}
\caption{
Ablation on the \textit{contrastive-branch masking ratio}.
All experiments keep the reconstruction-branch masking unchanged.
}
\label{tab:ablation_maskratio}
\end{table}


\begin{table}[!ht]
\centering
\setlength{\tabcolsep}{6.5pt}
\resizebox{\linewidth}{!}{
\begin{tabular}{l ccc ccc c}
\toprule
\textbf{Distillation} 
& \multicolumn{3}{c}{\textbf{VGGSound (V$\rightarrow$A)}} 
& \multicolumn{3}{c}{\textbf{VGGSound (A$\rightarrow$V)}} 
& \textbf{Classification} \\
\cmidrule(lr){2-4} \cmidrule(lr){5-7} \cmidrule(lr){8-8}
& R@1 & R@5 & R@10 & R@1 & R@5 & R@10 & AS20K (mAP) \\
\midrule
w/o & 29.5 & \textbf{54.7} & 63.9 & 29.1 & 53.1 & 63.5 & 30.5 \\
\rowcolor{gray!15}
w/  & \textbf{31.3} & 54.4 & \textbf{63.9} & \textbf{30.3} & \textbf{53.1} & \textbf{64.3} & \textbf{32.0} \\
\bottomrule
\end{tabular}
}
\caption{
Ablation on the \textit{teacher–student distillation} mechanism.
All experiments use a 0.50 masking ratio in the contrastive branch.
}
\label{tab:ablation_teacher}
\end{table}


\begin{table}[!ht]
\centering
\setlength{\tabcolsep}{6.5pt}
\resizebox{\linewidth}{!}{
\begin{tabular}{l ccc ccc c}
\toprule
\textbf{Masking Strategy} 
& \multicolumn{3}{c}{\textbf{VGGSound (V$\rightarrow$A)}} 
& \multicolumn{3}{c}{\textbf{VGGSound (A$\rightarrow$V)}} 
& \textbf{Classification} \\
\cmidrule(lr){2-4} \cmidrule(lr){5-7} \cmidrule(lr){8-8}
& R@1 & R@5 & R@10 & R@1 & R@5 & R@10 & AS20K (mAP) \\
\midrule
Random Mask & 31.2 & 54.3 & 64.8 & 30.3 & 54.5 & \textbf{65.7} & 30.2 \\
\rowcolor{gray!15}
Distinct Guided Mask & \textbf{31.3} & 54.4 & 63.9 & 30.3 & 53.1 & 64.3 & \textbf{32.0} \\
Probabilistic Guided Mask  & 29.6 & \textbf{56.1} & \textbf{65.4} & \textbf{30.5} & \textbf{54.9} & 65.0 & 29.8 \\
\bottomrule
\end{tabular}
}
\caption{
Ablation on the \textit{teacher-guided masking} strategy.
All experiments are conducted using a 0.50 masking ratio in the contrastive branch.
}
\label{tab:ablation_guidedmask}
\end{table}

%% file: 5_Conclusion.tex
\section{Conclusion}
In this work, we introduced a teacher-guided dual-path framework for audio-visual representation learning, revisiting how masked reconstruction and contrastive alignment should interact in CAV-MAE-style models. We argued that conventional joint optimization is limited by the use of a shared masked view for both objectives, which can weaken the contrastive branch through objective coupling and incomplete semantic context. To address this issue, we proposed three complementary designs: (1) a dual-pass training scheme that enables reconstruction and contrastive learning to operate on separate views; (2) a lightweight teacher-student pipeline that provides an auxiliary full-view consistency signal for masked global embeddings; and (3) a teacher-guided masking strategy that constructs a more structured contrastive view.

Our results show that objective-specific views are important for balancing reconstruction and alignment in audio-visual pretraining. More broadly, this work suggests that combining asymmetric masking with lightweight teacher guidance provides a simple and effective framework for improving large-scale audio-visual representation learning.

%% file: 6_Acknowledgement.tex
\section*{Acknowledgments}
This work was supported by the National Key R\&D Program of China under Grant No.2023ZD0120400, in part by the National Natural Science Foundation of China (No.62472423, 62276260, and 62206283).